\documentclass[a4paper,11pt]{article}

\usepackage{amsmath,amssymb}
\usepackage{graphicx}        % standard LaTeX graphics tool
\usepackage[top=1.4in,left=1.1in,bottom=0.8in,right=0.8in]{geometry}
\usepackage{enumerate,enumitem}
\usepackage{bm,ragged2e,array}
\usepackage{color,booktabs,multirow}
\usepackage[authordate,backend=biber]{biblatex-chicago}
\AtBeginBibliography{\footnotesize}

\DeclareMathAlphabet{\mathpzc}{OT1}{pzc}{m}{it}

\DeclareFontFamily{OT1}{pzcbf}{}
\DeclareFontShape{OT1}{pzcbf}{m}{n}{<-> s * [1.100] pzcmi8t}{}
\DeclareMathAlphabet{\mathpzcbf}{OT1}{pzcbf}{m}{n}

\usepackage{xparse}  % smarter commands

\newcommand\code{\bgroup\@makeother\_\@makeother\~\@makeother\$\@codex}
\def\@codex#1{{\normalfont\ttfamily\hyphenchar\font=-1 #1}\egroup}

\DeclareFieldFormat{doi}{%
  \mkbibacro{DOI}\addcolon\space
  \ifhyperref
    {\href{https://doi.org/#1}{\nolinkurl{#1}}}
    {\nolinkurl{#1}}}
\def\phipar{\varphi}
\def\mupar{\mu}
\def\sigmapar{\sigma}
\def\rhopar{\rho}

\def\hpar{h}

\NewDocumentCommand{\Sigmay}{o}{\bm{\IfValueTF{#1}{#1}{}\Sigma}}  % usage: \Sigmay or \Sigmay[\tilde]  (tilde is also bold face then!)
\newcommand{\deffsvpars}[2]{%
  \expandafter\def\csname phi#1\endcsname{\phipar^{#2}}
  \expandafter\def\csname mu#1\endcsname{\mupar^{#2}}
  \expandafter\def\csname sigma#1\endcsname{\sigmapar^{#2}}
  \expandafter\def\csname rho#1\endcsname{\rhopar^{#2}}
  \expandafter\def\csname h#1\endcsname{\hpar^{#2}}
  \expandafter\def\csname Sigma#1\endcsname{\Sigmay^{#2}}
}
\deffsvpars{idi}{\text{idi}}   % define all at once
\deffsvpars{fac}{\text{fac}} % define all at once
\newcommand{\Real}{\mathbb{R}}

\newcommand{\Student}[3]{t_{#1}(#2,#3)}   % \Student{df}{mean}{scale}
\newcommand{\Uniform}[2]{\mathcal{U}(#1,#2)}   % \Uniform{a}{b}
\newcommand{\Exponential}[1]{\mathcal{E}(#1)}   % \Exponential{a}
\NewDocumentCommand{\Gammadist}{o m m}{\mathcal{G}\IfValueTF{#1}{\left}{}({#2},{#3}\IfValueTF{#1}{\right}{})}   % \Gammadist{a}{b} or \Gammadist[1]{a}{b}
\NewDocumentCommand{\Normal}{o o m m}{\mathcal{N}\IfValueTF{#1}{_{#1}}{}\IfValueTF{#2}{\left}{}({#3},{#4}\IfValueTF{#2}{\right}{})}
\title{Bayesian Estimation of the Degrees of Freedom Parameter of the Student-$t$ Distribution---A Beneficial Re-parameterization}
\author{Darjus Hosszejni}%\textsuperscript{ORCID: 0000-0002-3803-691X}
\usepackage[hidelinks]{hyperref}

\begin{document}

\maketitle

\abstract{
	In this paper, conditional data augmentation (DA) is investigated for the degrees of freedom parameter $\nu$ of a Student-$t$ distribution.
	Based on a restricted version of the expected augmented Fisher information, it is conjectured that the ancillarity DA is progressively more efficient for MCMC estimation than the sufficiency DA as $\nu$ increases; with the break even point lying at as low as $\nu\approx4$.
	The claim is examined further and generalized through a large simulation study and a  application to U.S.\ macroeconomic time series.
Finally, the ancillarity-sufficiency interweaving strategy is empirically shown to combine the benefits of both DAs.
The proposed algorithm may set a new standard for estimating $\nu$ as part of any model.}

\section{Introduction} \label{sec:intro}

The Student-$t$ distribution is a standard item in the statistician's toolbox.
Utilized as the regression error distribution, it robustifies parameter estimation against extreme observations in the data set~\parencite{Gelman2013a}.
Furthermore, the properties of Student-$t$ are classical building blocks of hierarchical models for financial~\parencites{Bollerslev1987,Harvey1994} and polychotomous data~\parencite{Albert1993}, among others.

Maximum likelihood (ML) estimation of the degrees of freedom ($\nu$) parameter can be cumbersome due to the complexity of the likelihood function; \textcite{Lange1989} give an overview and develop a quasi-Newton, an expectation-maximization (EM), and a score function-based approach to ML.
\textcites{Liu1997} reviews extensions to the EM algorithm for the multivariate Student-$t$ distribution.
In the non-asymptotic branch of frequentist literature, \textcite{Fraser1976} discusses structural inference for the parameters, and \textcites{Sutradhar1986,Singh1988} provide method of moments estimators for $\nu$.

Bayesian treatments of robust regression date back to \textcite{DeFinetti1961}, who explores scale-mixtures of normal distributions as a tool for handling outliers and presents the Cauchy distribution as an example, which corresponds to $\nu=1$.
Still with a fixed $\nu$, \textcite{Zellner1976} discusses posterior inference for the parameters of a multivariate linear Student-$t$ regression using conjugate prior distributions.
\textcites{Ramsay1980,West1984} further extend de Finetti's framework and provide more examples including the Student-$t$ distribution.
\textcites{Chib1991} demonstrate conditions, when this type of robust regression can not work due to a non-informative likelihood for $\nu$.

In terms of Bayesian inference for $\nu$ using Markov chain Monte Carlo methods~\parencite[MCMC;][]{Tierney1994}, the Gibbs sampler~\parencite{Geman1984} by \textcite{Geweke1993} has been established as the standard method and it comprises the main focus in this paper.
The Gibbs sampler relies on the scale-mixture representation of the Student-$t$ distribution to draw a posterior sample using a model with data augmentation~\parencite[DA;][]{VanDyk2001}.
There exist many options for DA; however, to our knowledge, only the one outlined by \textcite{Geweke1993}, termed sufficient augmentation (SA), has been in use for Student-$t$.

Multiple examples demonstrate that SA may not work well for sampling $\nu$; see, for instance, \textcite[efficiency values in Appendix C]{Geweke1992}, or \textcite{Nakajima2009a}, who blame the hierarchical structure and the lack of marginalization for the low sampling efficiency.
However, inspired by the general framework of \textcite{Papaspiliopoulos2007}, an alternative way to higher efficiency is found.
They demonstrate through multiple example models that the transformation of SA to the ancillary augmentation (AA) often boosts the efficiency already quite drastically; unfortunately, whether AA or SA is superior depends on the data set.
\textcite{Papaspiliopoulos2008} propose as the way out of this data-dependence a hybrid Gibbs-sampler, which randomly picks AA or SA in every pass; they conclude that a combination of sufficiency and ancillarity may be desirable in any setting.
Finally, \textcite{Yu2011} introduce the ancillarity-sufficiency interweaving strategy (ASIS), which exploits the complementary roles of AA and SA in any hierarchical model and integrates them into one generally effective recipe.
The applicability and utility of ASIS is documented in the literature~\parencite[see, e.g.,][]{Kastner2014,Bitto2019,Monterrubio-Gomez2020}.

The contribution of this paper is two-fold: the ancillary parameterization for the Student-$t$ distribution is introduced, and it is shown through simulations that the AA sampler explores the posterior space more efficiently than the usual SA sampler for a large set of data generating processes; furthermore, an implementation of ASIS is proposed that can be applied as a plug-and-play substitute in any MCMC procedure for models that include the Student-$t$ distribution with unknown degrees of freedom.

The remaining part of the paper is structured as follows:
Section~\ref{sec:model} introduces the model and the notation. Section~\ref{sec:alg} provides details on measurement of efficiency,  the Gibbs sampling algorithm, and the corresponding conditional distributions.
Section~\ref{sec:sim} showcases the efficiency of the introduced method on a large grid of simulated data sets, and Section~\ref{sec:app} is an application inspired by~\textcite{Geweke1994a}.
Section~\ref{sec:con} concludes.

\section{Simple Student-$t$-Model}\label{sec:model}

%Following the notation by \textcite{Chib1991},
The simplest setup with Student-$t$ errors is investigated;
\begin{equation}\label{eq:baseline}
	y\sim\Student{\nu}{0}{1},
\end{equation}
where $\Student{\nu}{\mu}{\sigma^2}$ denotes the Student-$t$ distribution with $\nu$ degrees of freedom, location $\mu$, and variance $\sigma^2$;
the kernel of its density function is $(1+x^2/\nu)^{-(\nu+1)/2}$.
Taking this simple model enables a pure analysis of sampling $\nu$, decoupled from other effects present in larger models; furthermore, Equation~\eqref{eq:baseline} is an ingredient of all models employing the Student-$t$ distribution.
At the same time, the simplicity of Equation~\eqref{eq:baseline} is not restrictive either due to the modular nature of Gibbs-sampling: in case an econometric model is extended by Student-$t$ errors, its corresponding MCMC procedure shall merely include the additional steps for drawing $\nu$.
This modularity is demonstrated in Section~\ref{sec:app} through a more complex model.

As \textcite{Zellner1976} points out, the density function of the Student-$t$ distribution can be formulated as a scale mixture of normal distributions, which inspires the introduction of missing data $\tau$ and the data augmented representation
\begin{equation}\label{eq:sufficient}
\begin{split}
	y\sim &\; \Normal{0}{\tau}, \\
	\tau^{-1}\sim &\; \Gammadist{\nu/2}{\nu/2},
\end{split}
\end{equation}
where $\Normal{\mu}{\sigma^2}$ denotes the univariate normal distribution with mean $\mu$ and variance $\sigma^2$, and $\Gammadist{\alpha}{\beta}$ is the gamma distribution with shape $\alpha$ and rate $\beta$.
\textcite{Papaspiliopoulos2007} call Equation~\eqref{eq:sufficient} the centered parameterization, while, following \textcite{Yu2011}, we prefer the term sufficient parameterization.
This terminology derives from the fact that $\tau$ is a sufficient statistic for $\nu$.

Based on Equation~\eqref{eq:sufficient}, it is possible to derive other data augmentation schemes, where the auxiliary variable and $\nu$ are a priori independent; these are called ancillary parameterization.
As \textcite{Papaspiliopoulos2007} point out, one such option is to replace $\tau$ by $u=F(\tau;\nu)$, where $F(x;\nu)$ is the cumulative distribution function (CDF) for the prior distribution of $\tau$.
Note that $u$ follows a standard uniform distribution a priori; therefore, the ancillary data augmentation for Equation~ \eqref{eq:baseline} is
\begin{equation}\label{eq:ancillary}
	\begin{split}
		y\sim &\; \Normal{0}{F^{-1}(u;\nu)}, \\
		u\sim &\; \Uniform{0}{1},
	\end{split}
\end{equation}
where $F^{-1}(u;\nu)$ is the inverse CDF for the prior distribution of $\tau$.

\begin{figure}[!ht]
	\centering
	\includegraphics[width=0.9\linewidth]{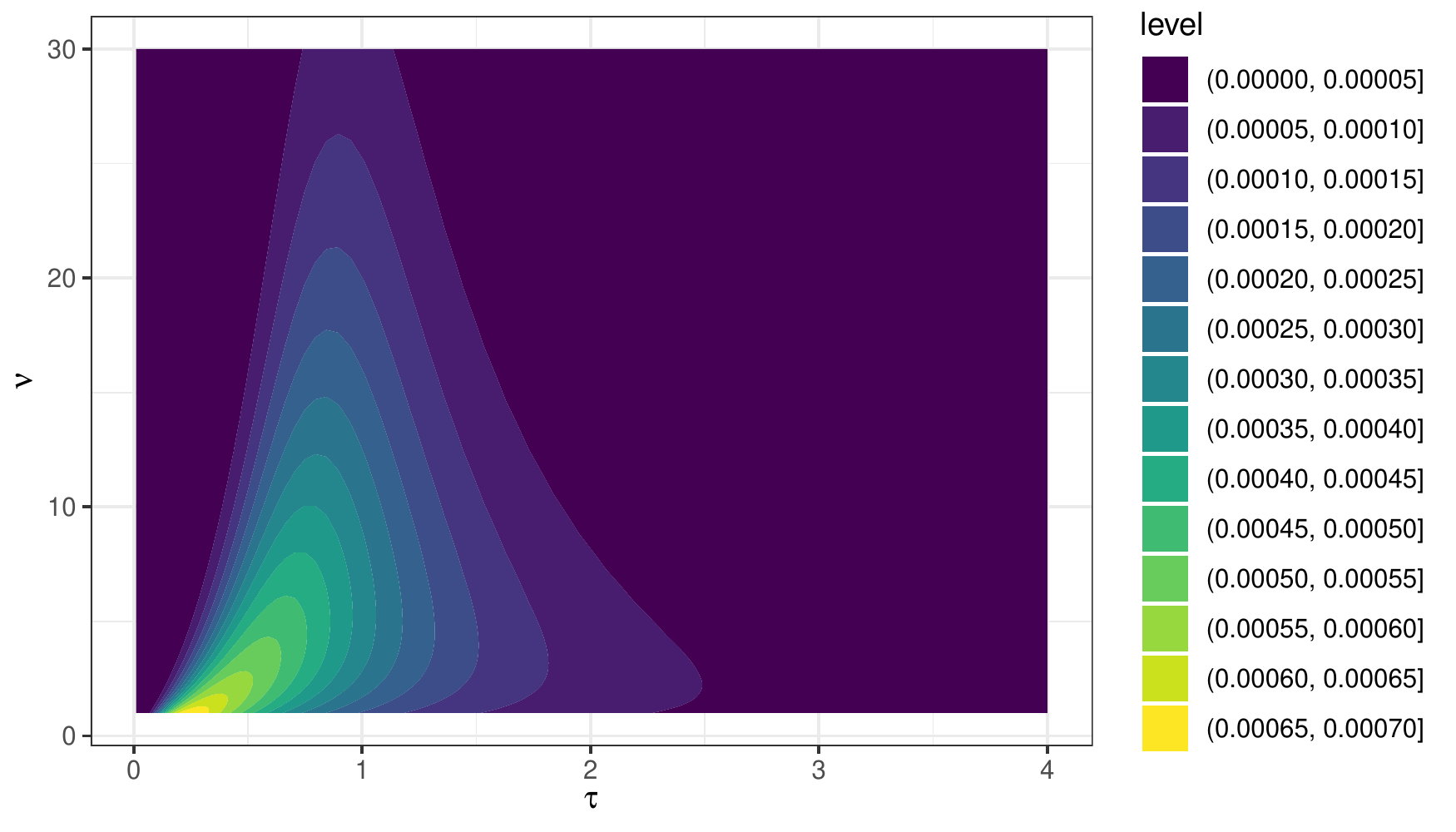}
	\includegraphics[width=0.9\linewidth]{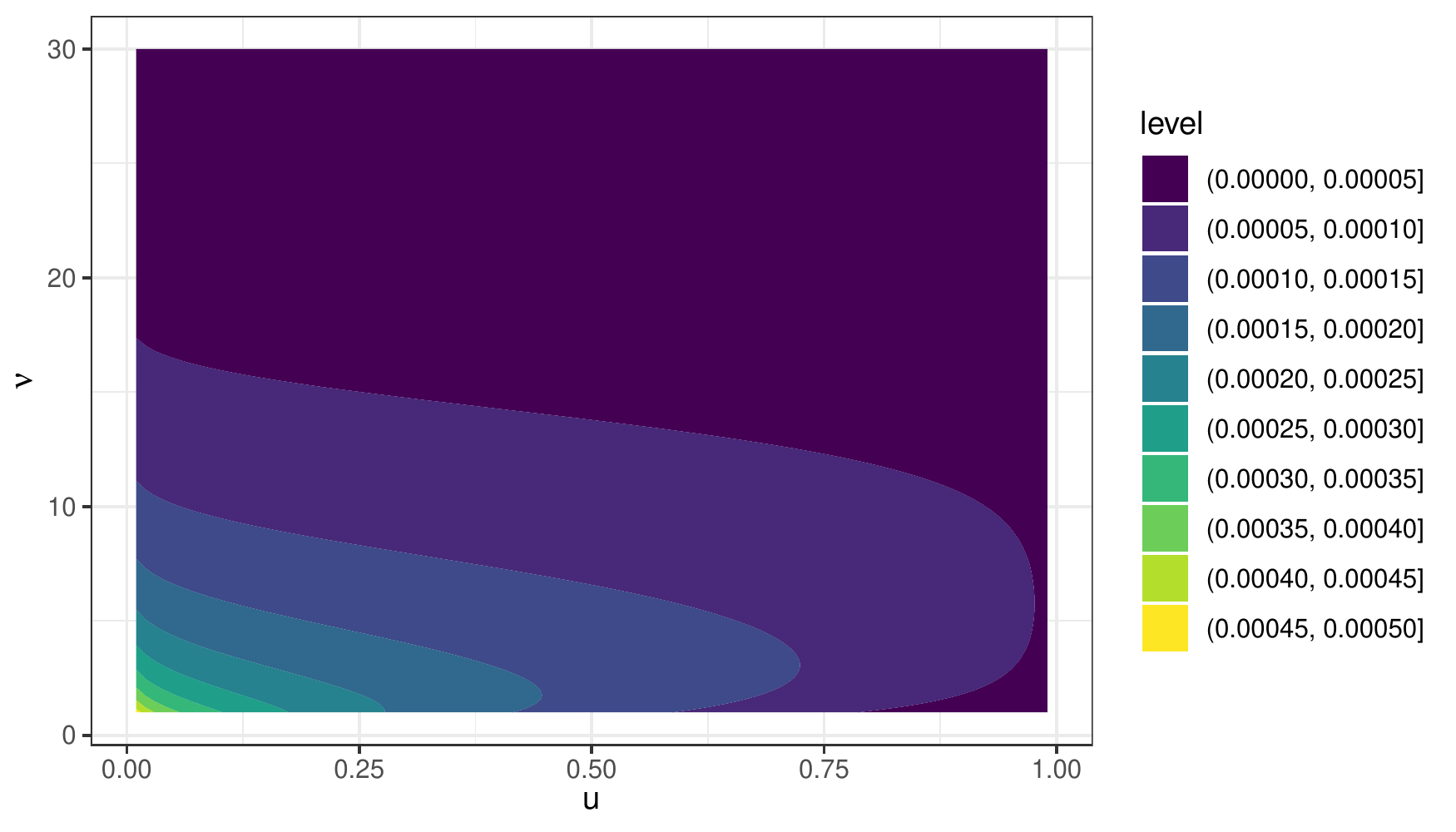}
	\caption{Contour plot of the bivariate density functions $p(\nu,\tau\mid y=0)$ (top) and $p(\nu,u\mid y=0)$ (bottom), where $\nu\in[1,30]$, $\tau\in(0,20]$, and $u\in(0,1)$.}
	\label{fig:y-small}
\end{figure}

\begin{figure}[!ht]
	\centering
	\includegraphics[width=0.9\linewidth]{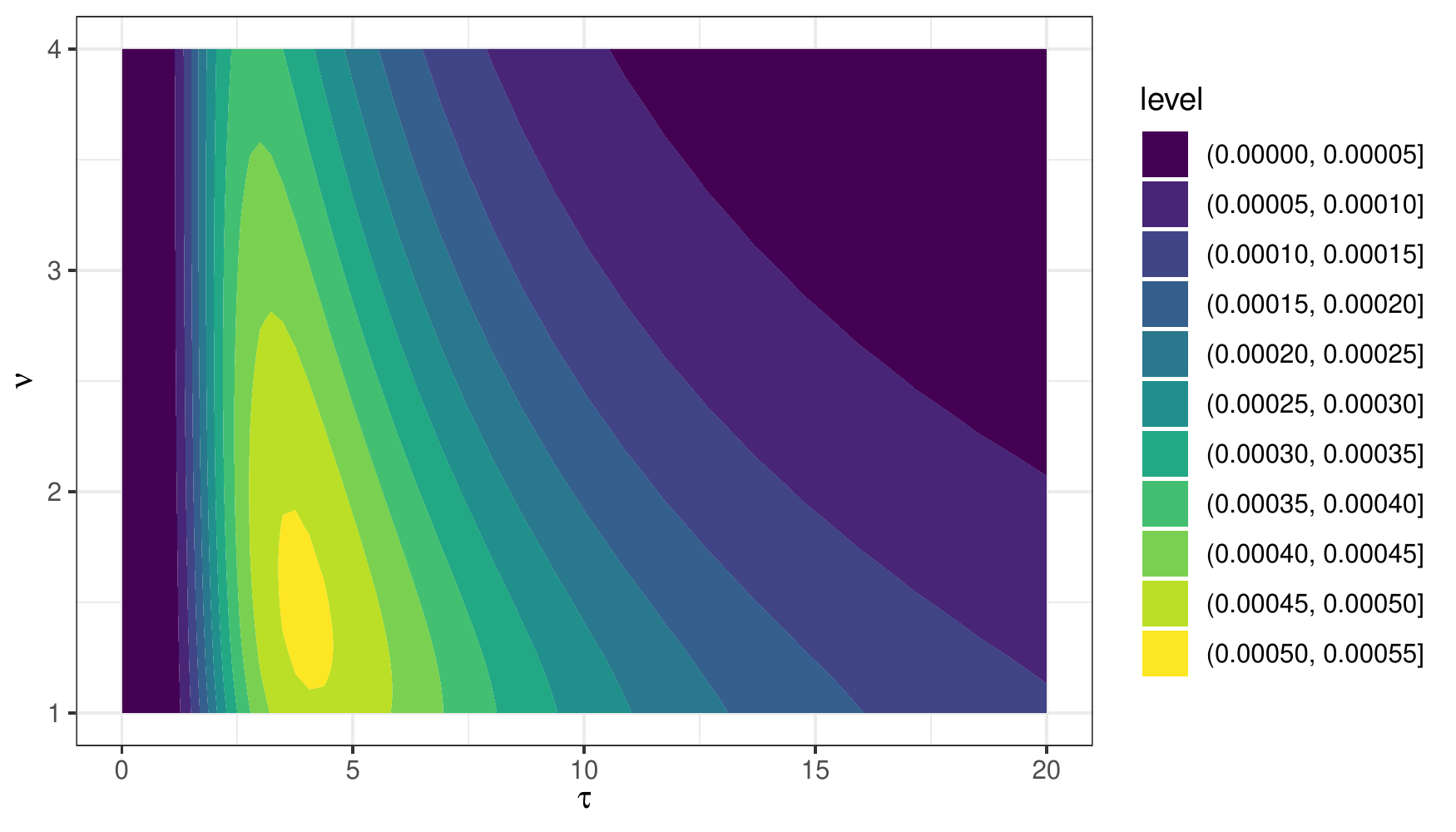}
	\includegraphics[width=0.9\linewidth]{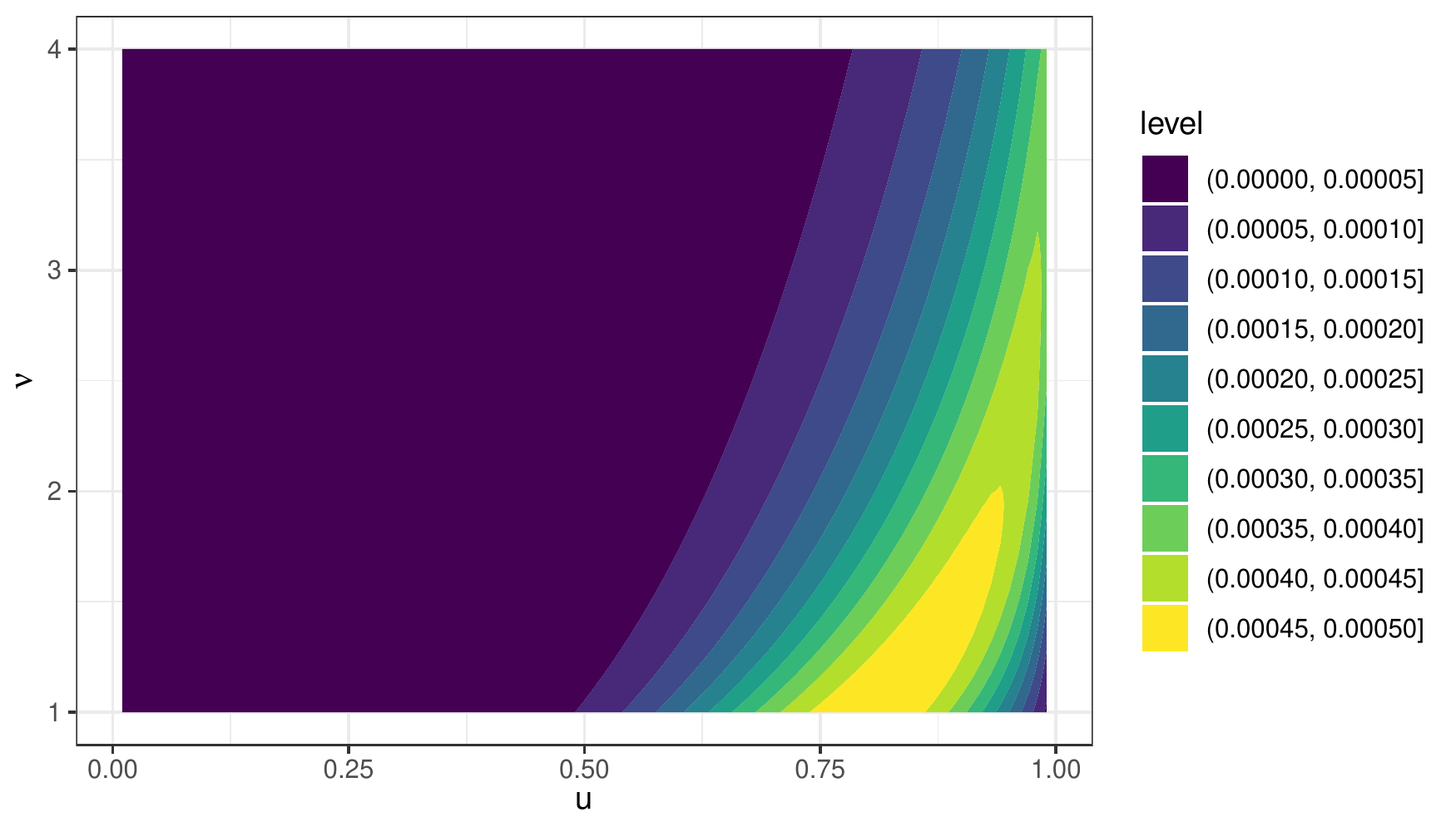}
	\caption{Contour plot of the bivariate density functions $p(\nu,\tau\mid y=4)$ (top) and $p(\nu,u\mid y=4)$ (bottom), where $\nu\in[1,30]$, $\tau\in(0,4]$, and $u\in(0,1)$.}
	\label{fig:y-large}
\end{figure}

As demonstrated by \textcite{Yu2011}, for improved sampling efficiency, it is beneficial to consider both a sufficient and an ancillary data augmentation scheme in the design of an MCMC sampling algorithm.
The added value in using both stems in the differences between the joint distributions $p(y,\nu,\tau)$ and $p(y,\nu,u)$; two slices for each distribution are depicted in Figures~\ref{fig:y-small} and~\ref{fig:y-large}, respectively.
The figures show contour plots of unnormalized\footnote{There is no closed form for these distributions, and normalization is done over the shown window, which contains most of the probability mass.} posterior densities $p(\nu,\tau\mid y=y_0)$ and $p(\nu,u\mid y=y_0)$ for small ($y_0=0$) and large ($y_0=4$) observation error.
The top panel of Figure~\ref{fig:y-small} depicts rotating shapes, which signals that the dependence between $\nu$ and $\tau$ is different at different parts of the space.
In comparison, this rotation is non-visible in the bottom panel, and, thus, dependence between $\nu$ and $u$ seems less complex; moreover, for the shown range, the shapes of isolines resemble concentric standard ellipses, which is a sign for approximate posterior independence between $\nu$ and $u$; a Gibbs-sampler can explore such a space very efficiently.
Figure~\ref{fig:y-large} shows an example, where a suspect outlier is modeled using Student-$t$ errors and a small value for $\nu$ is expected; here, the dependence between $\nu$ and $\tau$ varies less in space than the dependence between $\nu$ and $u$.
In conclusion, SA seems beneficial for fat tailed observations; on the other hand, AA may excel for lighter tailed observations.

The Bayesian description of the model is completed by the prior distribution $p(\nu)$.
Following \textcite{Geweke1993}, an exponential distribution is assumed;
\begin{equation}\label{eq:priornu}
	\nu\sim\Exponential{\lambda},
\end{equation}
where $\lambda$ denotes the rate parameter.
This choice is motivated by the availability of a suitable sampling procedure for SA using this prior~\parencite{Geweke1992}.
It is to be noted, that the adaptation of the proposed methods to alternative priors, such as the uniform distribution over an interval $[a,b]$~\parencite{Chib2002} or a gamma distribution~\parencite{Nakajima2012}, is uncomplicated.
This is discussed in Section~\ref{sec:mcmc}.

\section{Bayesian Estimation}\label{sec:alg}

\begin{figure}[!t]
	\centering
	\includegraphics[width=\linewidth]{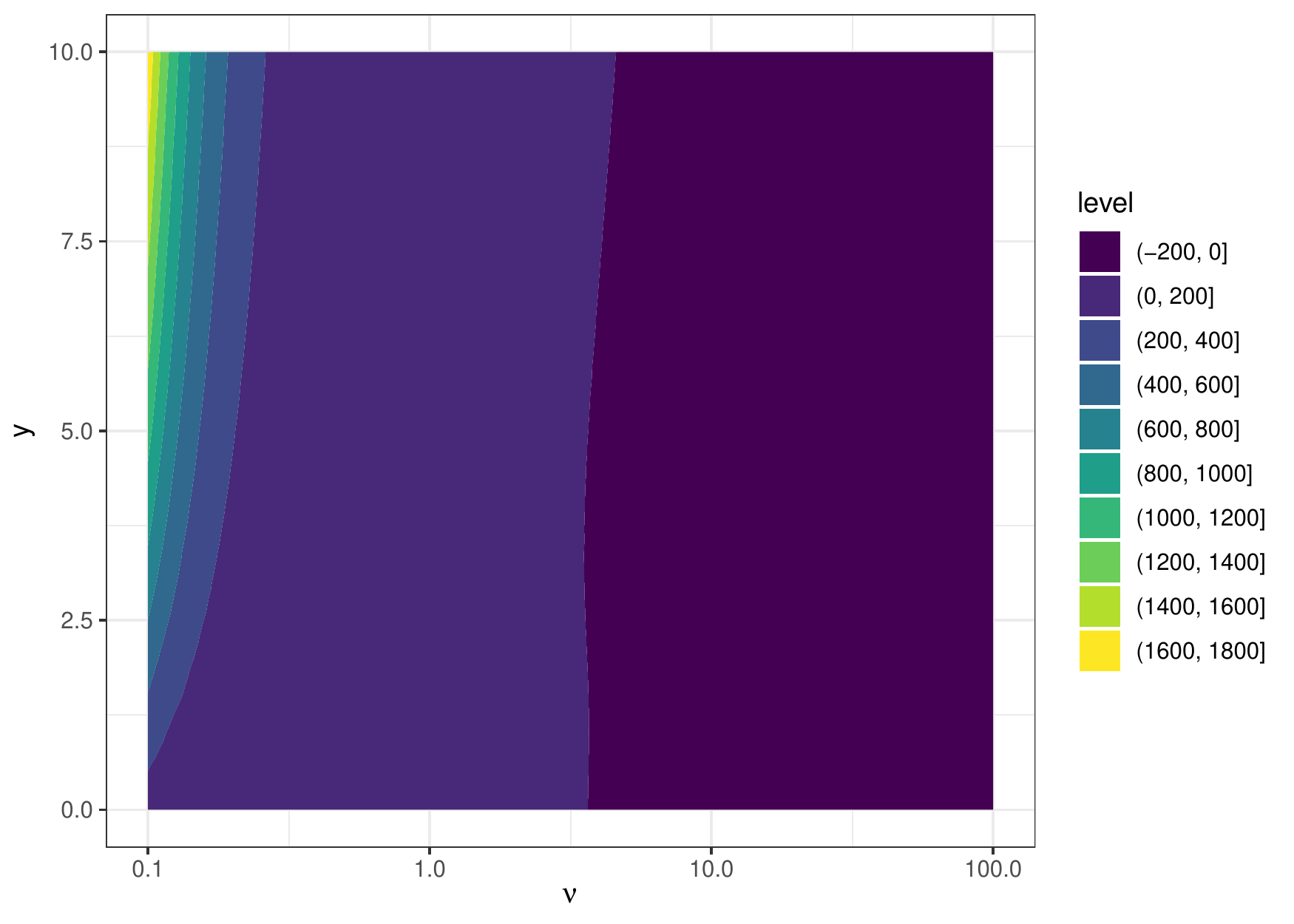}
	\caption{Contour plot of the estimated difference $I_u(y,\nu)-I_\tau(y,\nu)$ in the region $(y,\nu)^\top\in[0,10]\times[0.1,100]$.
	The depicted function is symmetric about $y=0$.
	Note that the horizontal axis is on a logarithmic scale.}
	\label{fig:fisher-contour}
\end{figure}

Two Gibbs-samplers are implemented that correspond to AA and SA, respectively; both of them consist of sequentially drawing random values from two conditional posterior distributions.
Above all, primary interest lies in how quickly the two resulting Markov chains mix, given different sets of observations; i.e., how close the posterior sample for $\nu$ is to a set of independent draws.
In particular, fast mixing is a desirable requirement for any MCMC procedure.
\textcite{VanDyk2001} discuss several measures for mixing and propose to minimize the expected augmented Fisher information matrix
\begin{equation*}
	\begin{split}
		I_\ast(y,\nu)=\text{E}\left[-\frac{\partial^2\log p(\nu\mid y,\ast)}{\partial\nu\cdot\partial\nu}\;\middle|\;y,\nu\right]
	\end{split}
\end{equation*}
for choosing between DAs, where $\ast\equiv\tau$ for SA and $\ast\equiv u$ for AA.
Although a simple derivation gives $I_\tau(y,\nu)=n(\Psi(\nu/2)-1/\nu)/2$, where $n$ denotes the number of observations and $\Psi(z)=\text{d}\log(\Gamma(x))/\text{d}z$ is the digamma function, we have been unsuccessful at computing $I_u$ analytically.
Fortunately, Monte Carlo (MC) estimation is possible.
Figure~\ref{fig:fisher-contour} exemplifies the difference between the estimated $I_u$ and $I_\tau$ for data sets of size 1.
The break even point (BEP) between AA and SA is at $\nu\approx4$ almost independently from $y$; that is where AA overtakes SA at this benchmark as $\nu$ increases.\footnote{For the evaluation of $I_u(y_0,\nu_0)$, a MC sample $\vec u_\text{MC}=(u^{(1)},\dots,u^{(L)})^\top$ of length $L=10000$ was simulated from $p(u\mid y=y_0,\nu=\nu_0)$ using the algorithm in Section~\ref{sec:mcmc}; next, $\partial^2\log p(\nu\mid y=y_0,u=u^{(i)})/\partial\nu\cdot\partial\nu$ is evaluated at points $i=1,\dots,L$; finally, the average of the values is taken to be the estimator for $I_u(y_0,\nu_0)$.
The numerical second derivative of $\log p(\nu\mid y,u)$ is approximated using Richardson's extrapolation with 6 steps, which is implemented in the \texttt{hessian} function from the software package \texttt{numDeriv}~\parencite{rnumderiv}.}%Importantly, this conclusion is robust against the MC error of this estimated image.

The comparison of $I_\tau$ and $I_u$ through MC quickly becomes impractical for larger data sets.
Therefore, an extensive computer simulation is conducted and empirical measurement of mixing is computed.
Ideally, the latter is based on exact sampling from the four conditional posterior distributions.

\subsection{Conditional Posterior Distributions}

Given data points $\vec{y}=(y_1,\dots,y_n)^\top$, the specification for $\vec\tau=(\tau_1,\dots,\tau_n)^\top$ is conditionally conjugate~\parencite{Geweke1993}, and, therefore, sampling from $p(\vec\tau\mid\vec y,\nu)$ is straightforward.
Next, as \textcite{Papaspiliopoulos2007} mention, sampling from $p(\vec u\mid\vec y,\nu)$ can be reduced to sampling $\vec\tau$.
%Nevertheless, no direct sampling strategy is known for either $p(\nu\mid\vec{y},\vec{u})$ or $p(\nu\mid\vec{y},\vec{\tau})$.

The rejection sampling (RS) method by \textcite{Geweke1993} is employed for $p(\nu\mid\vec{y},\vec{\tau})$: denote by $\xi^\ast$ the unique solution to
\begin{equation*}
	(\log(\xi/2)+1-\Psi(\xi/2))n/2+\xi^{-1}-\eta=0,
\end{equation*}
where $\eta=0.5\sum_{i=1}^{n}(\log(\tau_i)+1/\tau_i)-\lambda$ is a sufficient statistic.
By finding $\xi^\ast$, the normalizing constant in $p(\nu\mid\vec{y},\vec{\tau})$ is estimated indirectly, which is necessary for RS.
Then, a proposal $\nu_\text{p}$ is drawn from an exponential distribution with rate $1/\xi^\ast$, and it is retained with probability
\begin{equation*}
	\exp(\eta\xi^\ast-1)\left(\frac{(\xi^\ast/2)^{\xi^\ast/2}}{\Gamma(\xi^\ast/2)}\right)^{-n}\exp(\nu_\text{p}(1/\xi^\ast-\eta))\left(\frac{(\nu_\text{p}/2)^{\nu_\text{p}/2}}{\Gamma(\nu_\text{p}/2)}\right)^n.
\end{equation*}

Last, we are unaware of exact sampling procedures for $p(\nu\mid\vec{y},\vec{u})$.
A modern and easy-to-implement substitute is to employ adaptive Metropolis~\parencite[AM;][]{Haario2001} within Gibbs, which can be used to sample from any unnormalized density that can be evaluated.
However, contrary to RS, AM produces an auto-correlated Markov chain, and therefore it is inferior to RS in our setup.
As a simple improvement, independent sampling is approximated by repeating AM in one step of the Gibbs-sampler.
Our implementation of AM proposes $\log(\nu_\text{p})$ for the logarithm of $\nu$ through a Gaussian random walk with variance $\sigma^2$.
Next, $\nu_\text{p}$ is accepted or rejected according to the Metropolis ratio $\alpha$.
The proposal distribution is tuned to achieve the famous $\alpha=0.44$~\parencite{Gelman1994,Gelman1997,Roberts2001} by increasing or decreasing $\sigma$ after batches of several hundred draws; this is similar to the examples by \textcite{Roberts2009}.

\subsection{MCMC Algorithm}\label{sec:mcmc}

Following \textcite{Geweke1993}, Equation~\eqref{eq:sufficient} naturally translates into the following MCMC scheme, henceforth termed algorithm SA, which simulates from the posterior distribution $p(\nu,\vec\tau\mid\vec{y})$:
\begin{enumerate}
	\item Initialize $\nu$;
	\item Draw $\tau_i^{-1}\sim\Gammadist[1]{\frac{\nu+1}2}{\frac{\nu+y_i^2}2}$ independently for each $i=1,\dots,n$;
	\item Draw $\nu\sim p(\nu\mid \vec{y},\vec{\tau})$, where $\vec{\tau}=(\tau_1,\dots,\tau_n)^\top$ using RS;
	\item Repeat Steps 2 and 3.
\end{enumerate}
For alternative prior distributions, Step 3 can be replaced by AM as long as it is possible to evaluate the prior density function $p(\nu)$.
This change may affect the mixing speed for two reasons: autocorrelation is introduced into the Markov chain, and $I_\tau$ might be altered.

Next, the two novel algorithms are introduced.
Corresponding to Equation~\eqref{eq:ancillary}, algorithm AA samples from the posterior distribution $p(\nu,u\mid \vec{y})$;
\begin{enumerate}
\item Initialize $\nu$;
\item Draw $u_i\sim p(u_i\mid y_i,\nu)$ independently for each $i=1,\dots,n$ by first drawing $\tau_i^{-1}\sim\Gammadist[1]{\frac{\nu+1}2}{\frac{\nu+y_i^2}2}$ and then computing $u_i=F(\tau_i;\nu)$~\parencite[inspired by][]{Papaspiliopoulos2007};
\item Draw $\nu\sim p(\nu\mid\vec{y},\vec{u})$, where $\vec{u}=(u_1,\dots,u_n)^\top$ using AM; this step includes the evaluation of $F^{-1}(u_i;\nu)$; repeat this step $k_\text{AA}$ times;
\item Repeat Steps 2 and 3.
\end{enumerate}
This is algorithm is trivially modified for other prior distributions: the Metropolis ratio $\alpha$ needs to be adapted to the new prior density.

Finally, algorithm ASIS combines SA and AA:
\begin{enumerate}
	\item Initialize $\nu$;
	\item Do Steps 2 and 3 of SA;
	\item Compute $u_i=F(\tau_i;\nu)$ for $i=1,\dots,n$;
	\item Do Step 3 of AA;
	\item Repeat Steps 2 through 4.
\end{enumerate}
Notice that ASIS is not a sequence of SA and AA; $\vec u$ is not sampled separately.
Rather, focus lies on the differences between $p(\nu\mid\vec y,\vec\tau)$ and $p(\nu\mid\vec y,\vec u)$, which are exploited in a minimal setup.

Correctness of the implementations for AA, SA, and ASIS is confirmed using a variant on Geweke's test~\parencite{Geweke2004}.
Specifically, if a new vector of observations is simulated at the end of every round of the particular algorithm, then the sampling distribution of $\nu$ is equivalent to its prior; and this is verified visually with the help of quantile-quantile-plots.

\section{Simulation Study}\label{sec:sim}

In order to make a general comparison between the mixing of AA, SA, and ASIS, an extensive simulation study is conducted.
To this end, the algorithms in Section~\ref{sec:mcmc} are implemented in the \textsc{R} language~\parencite[version 4.1.0]{rlang2021}.
Then, multiple independent posterior Markov chains are generated for data simulated from Equation~\eqref{eq:baseline}; for AA, $k_\text{AA}=20$ is set.
Next, convergence of the Markov chains is automatically validated by ensuring that their $\hat R$ value~\parencite{Vehtari2021} is smaller than the recommended $1.1$; this is done using the \texttt{rhat} function of the R package \texttt{posterior}~\parencites{rposterior}.
Finally, relative numerical efficiencies~\parencite[RNE;][]{Geweke1989} of the chains are reported.
RNE is defined for an identically distributed but serially correlated Monte Carlo sample $\{\theta^{(j)}\}$ and a function of interest $g$ as $\hat{\text{var}}(g)/\hat{S}(0)$, where $\hat{\text{var}}(g)$ is a consistent estimator for the variance of the distribution $p(\theta)$ and $\hat{S}(\omega)$ is a consistent estimator for the spectral density $S(\omega)$ of $\{g(\theta^{(j)})\}$, $j=1,\dots,M$.%
\footnote{RNE is the reciprocal of the inefficiency factor reported in other works~\parencites{Nakajima2009a}.}
This computation is implemented in function \texttt{effectiveSize} from the \texttt{coda} package~\parencite{rcoda}.
%; only, $\text{RNE}\cdot n$ is returned.
%Since $n$ is known, recovering RNE is a trivial step.

\subsection{Setup of Data Generating Process}\label{sec:setup}

A grid of true values $\nu_\text{true}=1,1.5,2,2.5,3,4,5,10,20,50,100$ and data set lengths $n=1,3,10,30,100,300,1000,3000,10000$ is considered for the comparison.
This covers ``realistic'' values~\parencite[$\nu_\text{true}>2$;][]{Gelman2013a} and both problematic regions of low and high degrees of freedom.
Moreover, dependence on the size of the data set can be evaluated.
Then, for each of the 99 setups, five data sets are simulated from Equation~\eqref{eq:baseline}.
Furthermore, to examine prior sensitivity, values $\lambda=0.05, 0.1, 0.2, 0.5, 1$ are considered, which correspond to prior expectations of $1$ through $20$ for $\nu$.
In addition, following the recommendations for the computation of $\hat R$, four independent Markov chains are simulated with initial values $\nu=0.5, 2, 10, 100$ (these receive the same $\hat R$ during post-processing), for each of the three algorithms and each setup.
Overall, $29700$ samples of length $M=10000$ are drawn after a burn-in phase of length $1000$, and samples with $\hat R>1.1$ are removed.

\subsection{Posterior Estimates}\label{sec:est}

A sample of interval estimators for $\nu$ is presented in Table~\ref{tab:estimates} for $n=1000$ and $\lambda=0.2$.
With the exception of two cells, the estimates by AA, SA, and ASIS are similar for all values of $\nu_\text{true}$.
The twin of these exceptions is an example of a general caveat about AA: for a long vector of heavy-tailed observations, the numerical evaluation of $F^{-1}$ may fail within AA; especially for values $\nu$ at the outskirts of the posterior distribution.
The issue materializes as a constant estimator $(100,100)$ in Table~\ref{tab:estimates}.\footnote{
Our implementation for $F^{-1}$ is based on the built-in \textsc{R} function \texttt{qgamma}.}
Typically, the Markov chain is constant in these settings because the MH step always rejects; in all other settings, AA is reliable.
Consequently, by choosing (our implementation of) AA, one assumes a complex restriction on the prior distribution $p(\nu,\vec u)$.
Moreover, the three rows for $\nu_\text{true}=100$ provide two further insights.
First, obviously, even at sample size $n=1000$, most of the posterior mass is closer to the prior expectation of $1/\lambda=5$ than to 100; indeed, the prior choice is quite influential compared to the flat likelihood for light-tailed data.
Second, the variation in the estimates from SA is the largest in this row both in relative and in absolute terms; the $90\%$ quantile ranges between $37.9$ and $40.7$.
This sheds light on the outstandingly deteriorating effectiveness of SA as $\nu_\text{true}$ increases; this is discussed in Section~\ref{sec:eff}.
As a result, its finite sample properties make SA less reliable for estimating a large $\nu$.
Importantly, ASIS does not suffer from any of the aforementioned issues; indeed, it combines the best of both worlds.

\begin{table}[!t]
	
	\caption{Estimated 10th and 90th percentiles for $\nu$.
		The same data of length $n=1000$ is used inside each group of $\nu_\text{true}$;
		further, $\lambda=0.2$ for each entry.
		Sets of chains with $\hat R\ge1.1$ are marked.}
	\label{tab:estimates}
	\centering
	\begin{tabular}[t]{rlcccc}
		\toprule
		$\nu_\text{true}$ &  & Initial $\nu=0.5$ & 2 & 10 & 100\\
		\midrule
		& AA & $(0.894, 1.01)^\ast$ & $(0.895, 1.01)^\ast$ & $(0.893, 1.01)^\ast$ & $( 100,  100)^\ast$\\
		\cmidrule{2-6}
		& SA & $(0.893, 1.01)$ & $(0.894, 1.01)$ & $(0.893, 1.01)$ & $(0.895, 1.01)$\\
		\cmidrule{2-6}
		\multirow{-3}{*}{\raggedleft\arraybackslash 1} & ASIS & $(0.894, 1.01)$ & $(0.893, 1.01)$ & $(0.894, 1.01)$ & $(0.894, 1.01)$\\
		\cmidrule{1-6}
		& AA & $(1.74, 2.05)^\ast$ & $(1.75, 2.05)^\ast$ & $(1.75, 2.05)^\ast$ & $( 100,  100)^\ast$\\
		\cmidrule{2-6}
		& SA & $(1.75, 2.05)$ & $(1.74, 2.05)$ & $(1.75, 2.05)$ & $(1.74, 2.04)$\\
		\cmidrule{2-6}
		\multirow{-3}{*}{\raggedleft\arraybackslash 2} & ASIS & $(1.75, 2.05)$ & $(1.75, 2.05)$ & $(1.75, 2.05)$ & $(1.74, 2.04)$\\
		\cmidrule{1-6}
		& AA & $(3.99, 5.25)$ & $(3.98, 5.23)$ & $(3.99, 5.26)$ & $(3.98, 5.23)$\\
		\cmidrule{2-6}
		& SA & $(   4, 5.31)$ & $(3.94, 5.21)$ & $(3.98, 5.25)$ & $(3.99, 5.26)$\\
		\cmidrule{2-6}
		\multirow{-3}{*}{\raggedleft\arraybackslash 5} & ASIS & $(3.98, 5.23)$ & $(3.98, 5.24)$ & $(3.98, 5.24)$ & $(3.99, 5.24)$\\
		\cmidrule{1-6}
		& AA & $(  10, 19.4)$ & $(10.2, 19.1)$ & $(10.2, 19.3)$ & $(10.2, 19.4)$\\
		\cmidrule{2-6}
		& SA & $(10.2, 19.4)$ & $(10.2, 19.4)$ & $(10.4, 19.8)$ & $(10.2, 18.6)$\\
		\cmidrule{2-6}
		\multirow{-3}{*}{\raggedleft\arraybackslash 20} & ASIS & $(10.2, 19.3)$ & $(10.1, 19.1)$ & $(10.1, 19.3)$ & $(10.2, 19.1)$\\
		\cmidrule{1-6}
		& AA & $(16.6, 39.9)$ & $(16.5, 40.1)$ & $(16.7, 39.9)$ & $(16.7, 40.2)$\\
		\cmidrule{2-6}
		& SA & $(16.7, 40.7)$ & $(16.8, 37.9)$ & $(17.1, 38.4)$ & $(  17, 38.9)$\\
		\cmidrule{2-6}
		\multirow{-3}{*}{\raggedleft\arraybackslash 100} & ASIS & $(16.6, 39.9)$ & $(16.7, 40.1)$ & $(16.7, 39.8)$ & $(16.7, 40.2)$\\
		\bottomrule
	\end{tabular}
\end{table}

\subsection{Sampling Efficiency}\label{sec:eff}

Detailed results are presented in Table~\ref{tab:numbers}.
In most cases of practical interest, i.e., where the observations have finite variance by assumption, the RNE of AA is at least comparable to that of the classical SA; and, for lighter-tailed inputs, AA is incomparably more efficient.
Nevertheless, as discussed in Section~\ref{sec:est}, AA may fail; these cases are marked as missing values in Table~\ref{tab:numbers}.
In particular, AA breaks down for the combination of a data set of length $n\gtrapprox100$ simulated with $\nu_\text{true}\lessapprox3$, and a chain with initialization $\nu=100$; however, apparent from the results, lighter-tailed data sets may be increasingly affected as their size grows.
On the other hand, AA is reliable in all other settings.
Luckily, ASIS proves to be a solution in this regard, as SA does not suffer from the issue; therefore, the chain can move out of this initial bad region.
Moreover, ASIS outperforms both AA and SA in all setups.
Even an RNE of 100\% is reached for the small $n=10$, which is measurably on par with an i.i.d.\ sample.

\begin{table}[!t]
	
	\caption{Mean RNE (\%) for AA, SA, and ASIS, and different values of $n$ and $\nu_\text{true}$.
	}\label{tab:numbers}
	\centering
	\begin{tabular}[t]{lrrrrrrrrrrrr}
		\toprule
		&  & $\nu_\text{true}=1$ & 1.5 & 2 & 2.5 & 3 & 4 & 5 & 10 & 20 & 50 & 100\\
		\midrule
		& $n=10$ & 23.2 & 39.6 & 53.1 & 63.9 & 70.2 & 77.6 & 81.7 & 87.7 & 90.1 & 90.9 & 91.1\\
		\cmidrule{2-13}
		& 100 & 4.3 & 15.4 & 22.0 & 30.5 & 38.2 & 45.5 & 51.6 & 60.4 & 68.0 & 70.9 & 71.3\\
		\cmidrule{2-13}
		& 1000 & - & - & 23.3 & 29.9 & 36.3 & 43.7 & 48.1 & 45.1 & 39.3 & 44.4 & 46.6\\
		\cmidrule{2-13}
		\multirow{-4}{*}{\raggedright\arraybackslash AA} & 10000 & - & - & - & - & 37.3 & 43.2 & 48.9 & 45.3 & 28.4 & 21.8 & 24.5\\
		\cmidrule{1-13}
		& 10 & 45.5 & 31.8 & 25.7 & 22.5 & 21.2 & 19.8 & 19.5 & 19.2 & 19.7 & 19.8 & 19.8\\
		\cmidrule{2-13}
		& 100 & 45.4 & 32.7 & 24.7 & 19.2 & 15.4 & 10.0 & 7.2 & 4.4 & 4.0 & 3.9 & 3.9\\
		\cmidrule{2-13}
		& 1000 & 44.4 & 31.3 & 23.2 & 17.8 & 14.2 & 9.5 & 6.8 & 2.2 & 1.0 & 0.9 & 0.9\\
		\cmidrule{2-13}
		\multirow{-4}{*}{\raggedright\arraybackslash SA} & 10000 & 44.6 & 31.5 & 23.5 & 18.2 & 14.5 & 9.9 & 7.1 & 2.3 & 0.7 & 0.3 & 0.2\\
		\cmidrule{1-13}
		& 10 & 76.7 & 80.8 & 86.0 & 90.4 & 93.1 & 95.3 & 96.8 & 99.4 & 99.8 & 100.6 & 100.6\\
		\cmidrule{2-13}
		& 100 & 63.1 & 59.5 & 58.9 & 60.4 & 62.1 & 65.5 & 65.4 & 68.3 & 73.5 & 74.3 & 76.1\\
		\cmidrule{2-13}
		& 1000 & 63.0 & 60.3 & 60.6 & 61.6 & 63.1 & 64.1 & 64.1 & 49.6 & 40.9 & 45.3 & 48.0\\
		\cmidrule{2-13}
		\multirow{-4}{*}{\raggedright\arraybackslash ASIS} & 10000 & 62.5 & 59.7 & 60.5 & 62.0 & 63.3 & 65.3 & 64.4 & 50.0 & 29.3 & 21.9 & 25.1\\
		\bottomrule
	\end{tabular}
\end{table}

\begin{figure}[!t]
\centering
\includegraphics[width=\linewidth]{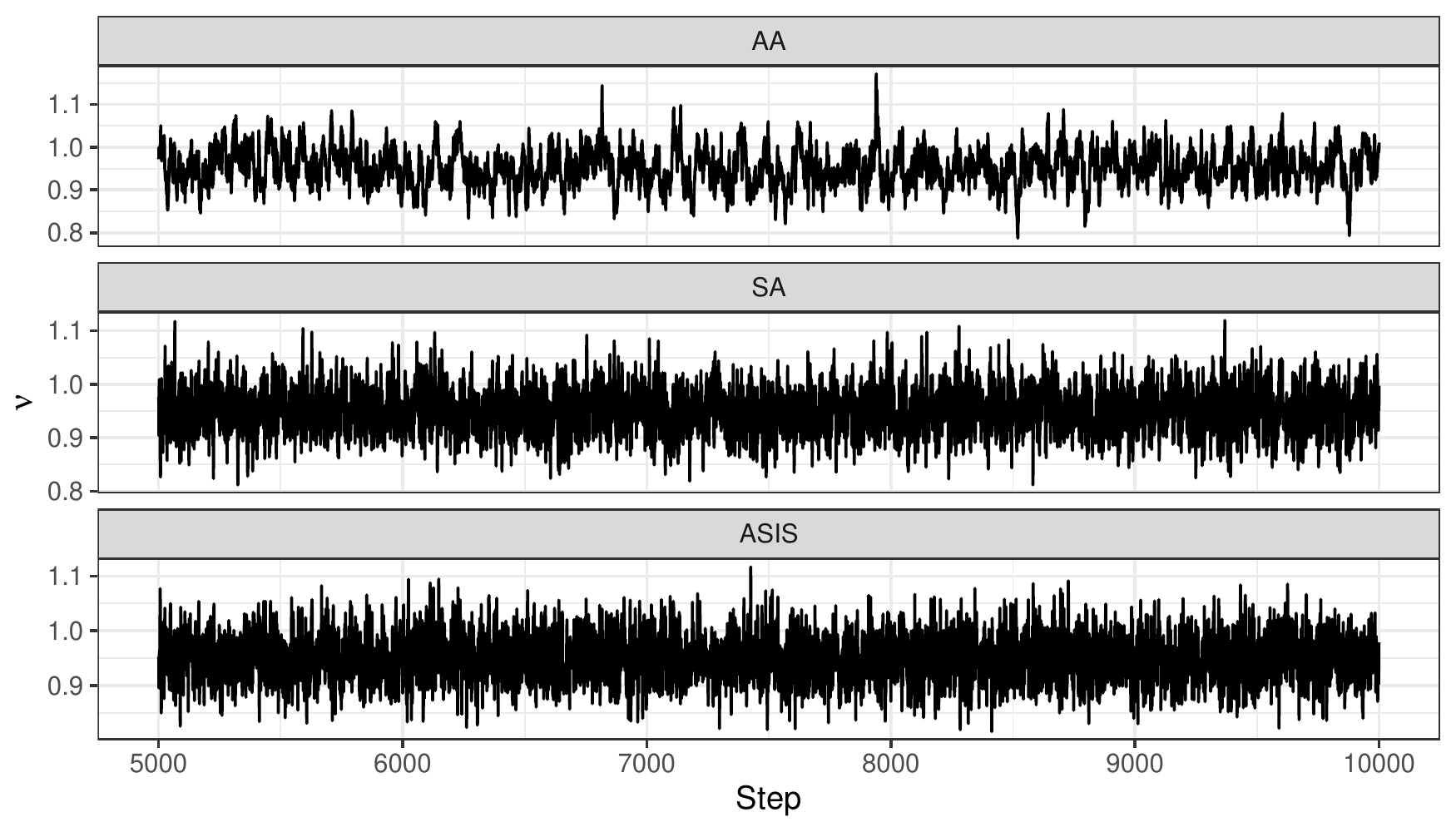}
\includegraphics[width=\linewidth]{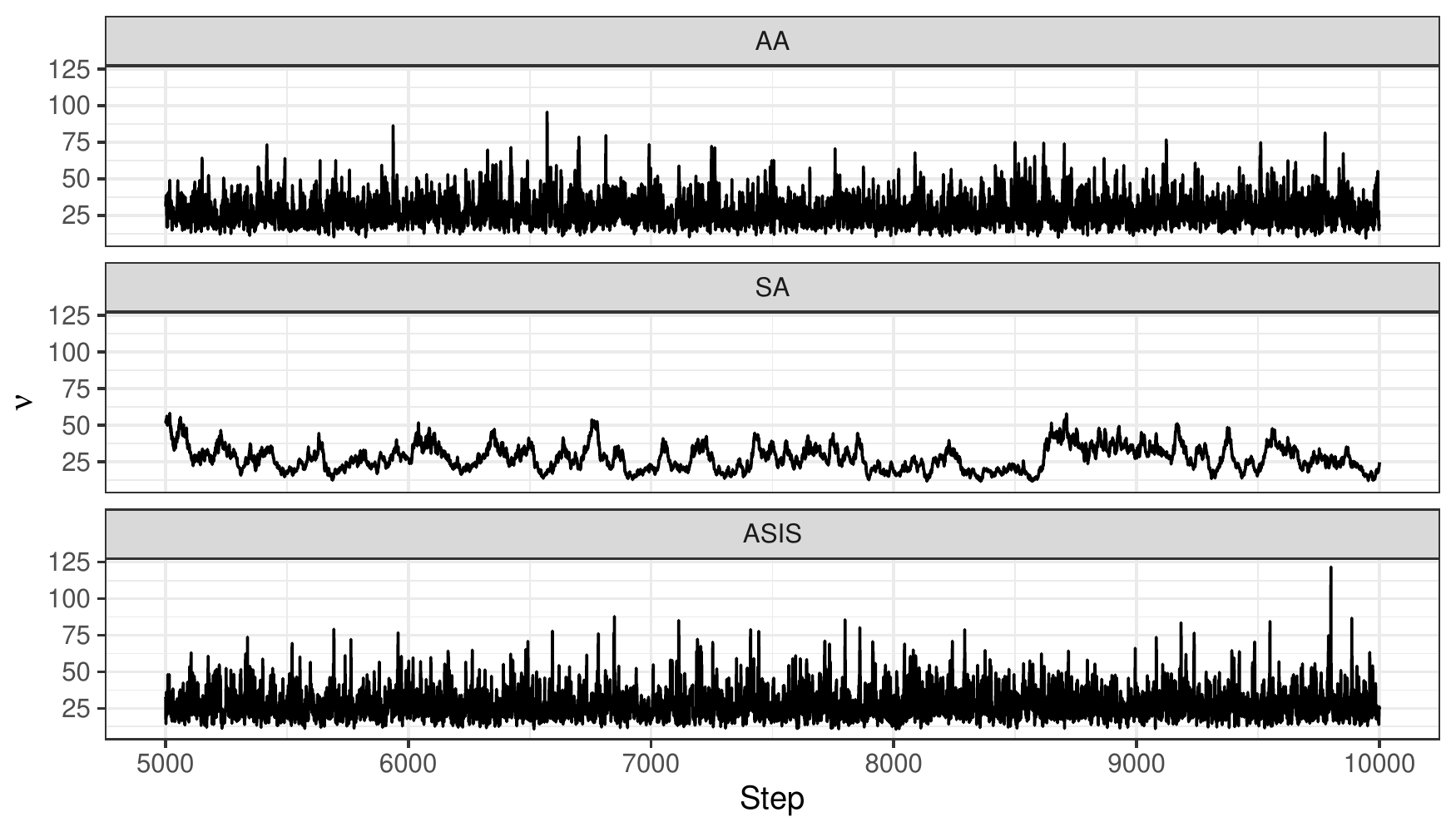}
\caption{Example trace plots for a posterior sample from $p(\nu\mid\vec y)$, where $n=1000$, $\lambda=0.2$, and the data $\vec y$ is simulated using $\nu=1$ (top) and $\nu=100$ (bottom).
For better visibility, only $5000$ draws are shown.}
\label{fig:example-trace}
\end{figure}

Some further results are highlighted using visualizations.
First, Figure~\ref{fig:example-trace} is a demonstration of the differences between AA, SA, and ASIS using trace plots: these are shown for hand-picked but representative examples.
The top three panels correspond to very heavy-tailed input data ($\nu_\text{true}=1$) and a good initialization for AA; SA and ASIS mix quickly, but AA seems to slightly lag behind.
In contrast, the bottom three panels represent the case of a light-tailed input vector ($\nu_\text{true}=100$); now, AA and ASIS show good performance, but SA is highly ineffective.

\begin{figure}[!t]
	\centering
	\includegraphics[width=\linewidth]{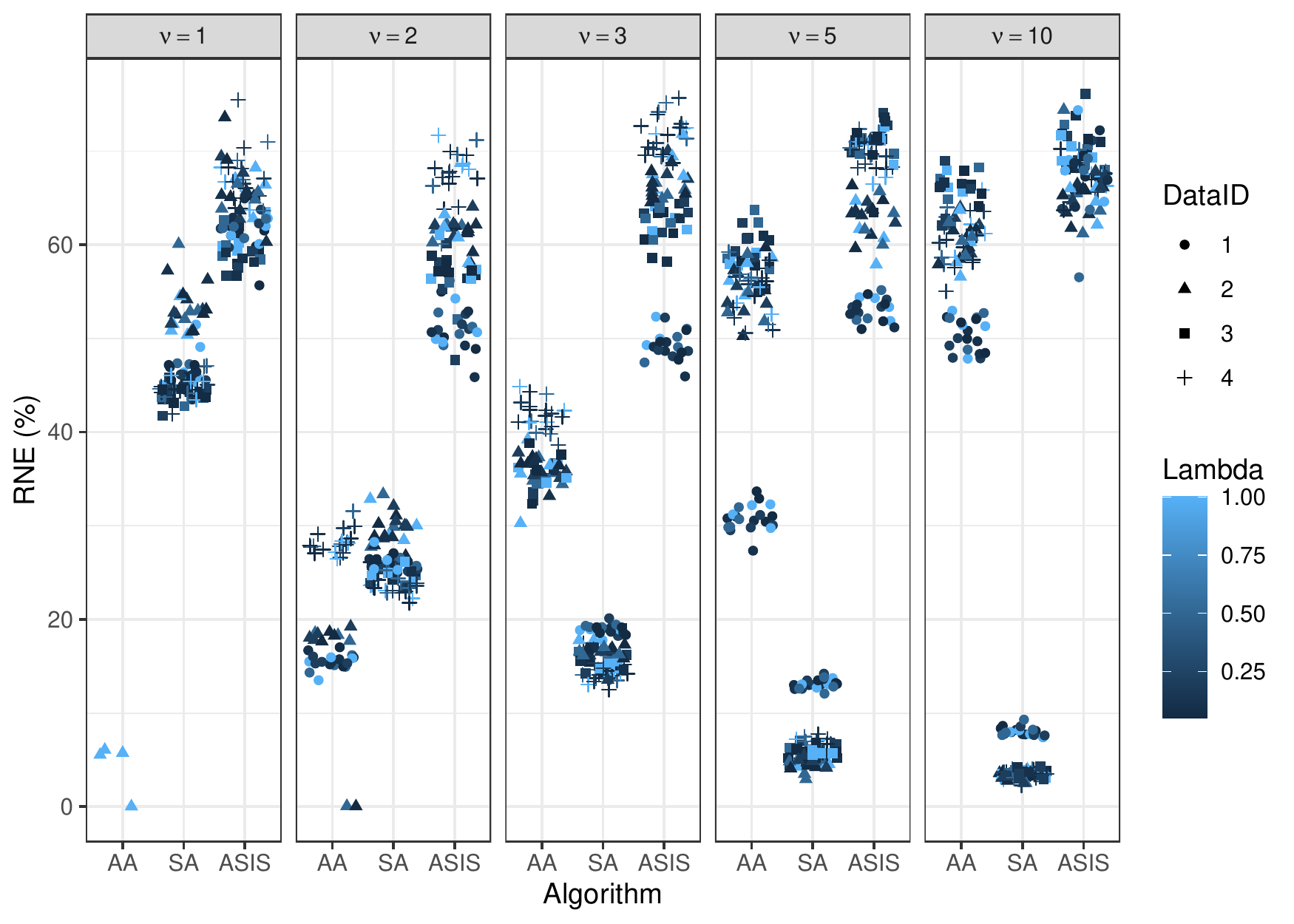}
	\caption{RNE in dependence on four variables: the algorithm (AA, SA, or ASIS), $\nu$, $\lambda$, and the input data set; $n=100$ is fixed.
	Points correspond to the individual Markov chains; therefore, four points belong to each setting.
	For increased visibility, only four out of the five simulated data sets are included, and horizontal jitter is applied to each point.
	Algorithm AA has only one chain for $\nu=1$ with an $\hat R<1.1$.}
	\label{fig:consistency}
\end{figure}

Next, Figure~\ref{fig:consistency} depicts a sample of RNE values conditional on the algorithm, $\nu_\text{true}$, $\lambda$, and the simulated data set; for improved readability, the plot is restricted to $n=100$, $\nu_\text{true}\in\{1,2,3,5,10\}$, and four out of five data sets.
Clearly, the sampling efficiency strongly depends on the exact properties of the observations.
For instance, the data set with $DataID=1$ is an outlier in the panel for $\nu=5$; and the reason for the comparably poor performance of AA is the extreme outlier $\simeq15$ contained in the data.
Similar behavior can be observed in the rightmost panel for $DataID=1$, which is due to the same random seed being used for generating data sets with the same $DataID$.
Further, at this sample size, $\lambda$ does not affect RNE.
Indeed, there is no clustering of colors in any of the 15 columns.
Since $\lambda$ cancels out in the Fisher information, this speaks for the similarity between RNE and $I_\tau$ or $I_u$.
In summary, part of the conclusion of Figure~\ref{fig:consistency} is consistent with that of Figure~\ref{fig:fisher-contour}: SA outperforms AA only for particularly heavy-tailed input data.
However, the data do affect the BEP, which, in addition, comes at a lower $\nu_\text{true}$ than the aforementioned value 4: between 2 and 3.
Finally and most importantly, ASIS is immune to both the numerical issues of AA and the quickly deteriorating efficiency of SA.

\begin{figure}[!t]
\centering
\includegraphics[width=\linewidth]{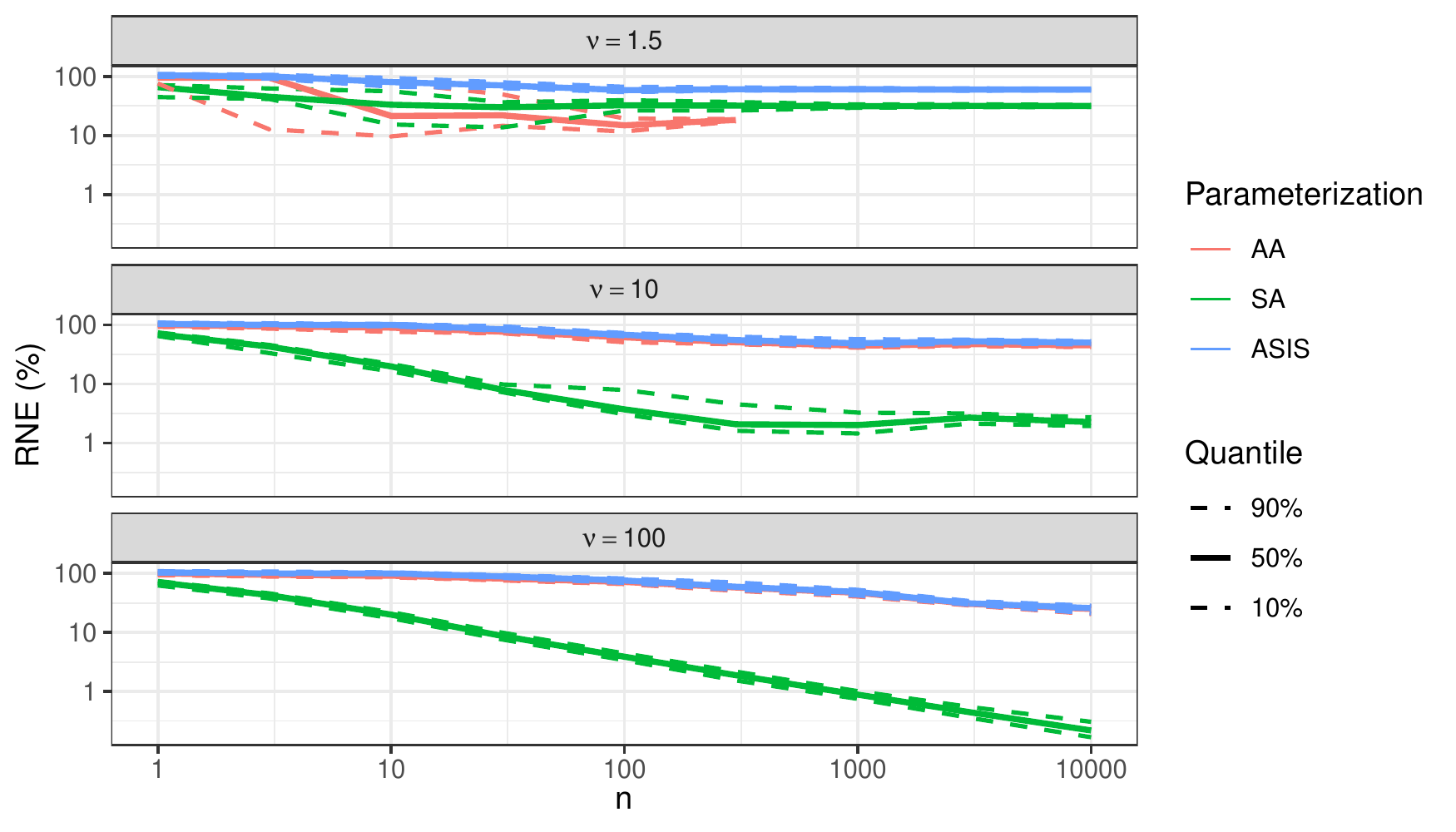}
\caption{RNE in dependence on $n$, $\nu_\text{true}$, and the algorithm.
	Quantiles are computed over the $100$ Markov chains that span over the five values for $\lambda$, four parallel chains, and five simulated data sets; as described in Section~\ref{sec:setup}.
	Both axes are on a logarithmic scale.
	Note that algorithm AA does not have chains for $n\ge1000$, whose $\hat R<1.1$.}
\label{fig:length-dependence}
\end{figure}

The significance of ASIS becomes even more apparent through Figure~\ref{fig:length-dependence}.
A summary of RNE is displayed for AA, SA, and ASIS, as it decreases with increasing $n$ and $\nu_\text{true}$.
Above all, the larger $\nu_\text{true}$ is, the greater the difference between the slopes for SA and ASIS.
Based on the bottom panel, one may hypothesize that the RNE of SA is linearly proportional to $1/n$ for light-tailed input; this should be connected to the multiplicative term $n$ in $I_\tau$.
Finally, for the last time, the dominance of ASIS is conspicuous.

\section{Application}\label{sec:app}

The efficiency of ASIS is evaluated on an extension of the 14 U.S.\ macroeconomic time series by \textcite{Nelson1982} as part of the model specified by \textcites{Geweke1992,Geweke1993,Geweke1994a}; these three closely connected publications are followed in this section. %; and, henceforth, for readability, only one is referred to.
%We modify the MCMC sampler of \textcite{Geweke1993} and closely follow his application, which was analyzed in depth by \textcite{Geweke1994a}.
%The data consist of real GNP, nominal GNP, real per capita GNP, industrial production index, total employment, total unemployment rate, GNP deflator, consumer price index (CPI), ...
The data set can be downloaded with the \textsc{R} package \texttt{tseries} \parencites{rlang2021,tseries2020}.%
\footnote{The data set was formerly stored in the data archive of the Journal of Business and Economic Statistics, currently at \url{http://korora.econ.yale.edu/phillips/data/np&enp.dat}.}
The observation series are annual, they start between 1860 and 1909, and all end in 1988.

\subsection{Model Specification}

The model is a re-parameterized auto-regressive linear model of lag five with a time trend and Student-$t$ increments, and, as highlighted by~\parencite{Geweke1993}, it is an alternative to the specifications employed by \textcite{Nelson1982} and others.
To underline the time-series nature of the data set, index $t$ is used in this section instead of $i$. Formally, observations $\{y_t\}$ for $t=1,\dots,T$ are assumed to be explained through
\begin{equation}\label{eq:geweke-large}
	\begin{split}
		y_t &= \gamma+\delta t+u_t, \\
		A^\star(L)u_t &= \varepsilon_t, \\
		A^\star(L) &= (1-\rho L)+(1-L)\sum_{j=1}^4a_jL^j, \\
		\{\varepsilon_t\} &\sim \text{i.i.d. }\Student{\nu}{0}{\sigma^2},
	\end{split}
\end{equation}
where $L$ is the lag operator, $\gamma\in\Real$ is the intercept, $\delta\in\Real$ is the slope of the main time trend, and the series $\{u_t\}$ stands for the cyclic deviations from the trend.
The cycle is assumed to vary according to the fifth degree polynomial specification for $A^\star$ with parameters $0\le\rho<1$ and $a_1,a_2,a_3,a_4\in\Real$.

It is possible to eliminate the lag operator \parencite{Geweke1993} and write down Equation~\eqref{eq:geweke-large} as
\begin{gather*}
	y_t = \gamma(1-\rho) + \delta\left(\rho-\sum_{j=1}^4a_j\right) + \delta(1-\rho)t +
	\rho y_{t-1} + \sum_{j=1}^4a_j(y_{t-j}-y_{t-j-1}) + \varepsilon_t.
\end{gather*}

Following~\textcite{Geweke1993}, the following prior distributions are assumed;
\begin{equation}\label{eq:prior1}
	\begin{split}
		p(\rho) &= 5\rho^4 I_{[0,1]}(\rho), \\
		\delta &\sim \Normal{0}{0.05^2}, \\
		a_j &\sim \Normal{0}{0.731\cdot 0.342^{j-1}}, \\
		\gamma &\sim \Normal{y_0}{10^2}, \\
		p(\sigma) &\propto \sigma^{-1}I_{(0,\infty)}(\sigma),
	\end{split}
\end{equation}
where $y_0$ is the first observation and it is not included in~\eqref{eq:geweke-large}, and $I_A(x)$ is the indicator function that takes $1$ for $x\in A$ and $0$ otherwise.
These prior distributions are in line with the hypotheses in the original work, where the model is employed with Gaussian increments ($\nu=\infty$) to show that business cycles deviate from a linear time trend in a non-stationary manner.
Finally, Equation~\eqref{eq:priornu} is adopted for $p(\nu)$ with $\lambda=0.333$.
%Finally, we adopt the prior for $\nu$ to be
%\begin{equation}
%	p(\nu)=\lambda\exp(-\lambda\nu)I_{(0,\infty)}(\nu),
%\end{equation}
%with hyperparameter $\lambda\in\Real^+$.

A variant on the algorithm by~\textcite{Geweke1993} is implemented;
\begin{enumerate}
	\item Initialize $(\gamma,\delta,\rho,a_1,a_2,a_3,a_4)^\top$ as the ordinary least squares (OLS) estimate (enforcing the bounds for $\rho$ if needed), $\sigma^2$ as the sample variance given the OLS estimates, $\tau\equiv1$, and $\nu=4$;
	\item Draw $(\gamma,\delta)^\top$ given all other variables from a bivariate normal distribution;
	\item Draw $(a_1,a_2,a_3,a_4)^\top$ given all other variables from a multivariate normal distribution;
	\item Draw $\rho$ given all other variables from a non-standard distribution using rejection sampling;
	\item Draw $\nu$ and $\vec\tau$ (for SA), $\vec u$ (for AA), or both (for ASIS) given all other variables as in Section~\ref{sec:mcmc}; note that $\{\varepsilon_t\}$ is a sufficient statistic;
%	\begin{enumerate}
%		\item Draw $\vec\tau$ given all other variables independently from an inverse-gamma distribution;
%		\item For the sufficient parameterization, draw $\nu$ given $\vec\tau$ using adaptive MH tuned for $23.4\%$ acceptance rate;
%		\item For the ancillary parameterization, compute $u_i=F(\tau_i;\nu)$, then draw $\nu$ given all other variables using adaptive MH tuned for $23.4\%$ acceptance rate, then compute $\tau_i=F^{-1}(u_i;\nu)$;
%		\item For ASIS, do steps 5.a, 5.b, and 5.c,
%	\end{enumerate}
	\item Draw $\sigma^2$ given all other variables from a scaled inverse-chi-square distribution with $T$ degrees of freedom.% $[\sum_{t=1}^T(\varepsilon_t^2/\tau_t)] / \chi^2_T$.
\end{enumerate}
Our modification only affects Step 5, where the method described in Section~\ref{sec:alg} is applied.
The exact distributions and methods of the remaining steps can be found in \textcites{Geweke1993} and in the software code as part of the Supplementary Material.

Similarly to Section~\ref{sec:mcmc}, correctness of our implementations for Steps 1-5 is ensured through Geweke's test~\parencite{Geweke2004}; the prior distributions are recovered by simulating data from Equation~\eqref{eq:geweke-large} at the end of each pass.
This does not work for Step 6 due to $p(\sigma)$ being improper.
There, the empirical quantile function of the conditional posterior distribution is compared to the theoretical quantile function.

\subsection{Posterior Estimates}

For all 14 variables, a posterior sample of size $10000$ is drawn after a burnin of $1000$ passes using AA, SA, and ASIS, resulting in 42 Markov chains; after the verification of convergence through trace plots, these numbers are considered to be adequate.
There were no numerical issues with the computation of $F^{-1}$ as part of AA.

Estimated posterior distributions are summarized in Table~\ref{tab:app-est}.
The three algorithms largely agree on the results, and there do not seem to be systematic biases present.
The figures suggest highly heavy-tailed posterior distributions for $\{\varepsilon_t\}$ in all cases, given the prior information.
Median values range between approximately $1.2$ and $6$, and, therefore, they fall on both sides of the previously measured BEP's.

\begin{table}[!t]
	
	\caption{Summary of posterior distributions for $\nu$.
	Credible intervals show the 10th and the 90th posterior percentiles after 10000 draws.}
	\label{tab:app-est}
	\centering
	\begin{tabular}[t]{lcccccc}
		\toprule
		 & \multicolumn{3}{c}{Median} & \multicolumn{3}{c}{80\% Credible Interval} \\
		 & AA & SA & ASIS & AA & SA & ASIS\\
		\midrule
		CPI & $2.21$ & $2.15$ & $2.17$ & $(1.57, 3.31)$ & $(1.55, 3.27)$ & $(1.54, 3.21)$\\
		Employment & $ 2.3$ & $2.34$ & $2.29$ & $(1.47, 4.03)$ & $( 1.5,    4)$ & $(1.49, 4.09)$\\
		GNP Deflator & $2.23$ & $2.21$ & $ 2.2$ & $(1.58, 3.27)$ & $(1.54, 3.25)$ & $(1.55, 3.26)$\\
		Industrial Production & $2.88$ & $2.81$ & $2.89$ & $(1.96, 4.71)$ & $(1.92, 4.49)$ & $(1.92, 4.74)$\\
		Interest Rate & $1.23$ & $1.22$ & $1.22$ & $(0.889, 1.72)$ & $(0.891, 1.73)$ & $(0.888, 1.75)$\\
		\addlinespace
		Money Stock & $3.55$ & $3.63$ & $3.56$ & $(2.26, 6.09)$ & $(2.27, 6.46)$ & $(2.27, 6.19)$\\
		Nominal GNP & $2.31$ & $2.32$ & $2.32$ & $(1.52, 3.79)$ & $(1.53, 3.75)$ & $(1.53, 3.77)$\\
		Real GNP & $   3$ & $3.09$ & $3.05$ & $( 1.8, 5.77)$ & $(1.82, 6.05)$ & $( 1.8, 5.85)$\\
		Real per Capita GNP & $2.79$ & $2.84$ & $2.87$ & $(1.68, 5.48)$ & $(1.65, 5.68)$ & $(1.69, 5.59)$\\
		Real Wages & $5.41$ & $5.41$ & $5.38$ & $(2.97, 10.4)$ & $(2.97, 10.3)$ & $(2.95, 10.2)$\\
		\addlinespace
		Stock Prices & $6.11$ & $5.99$ & $5.99$ & $(3.74, 10.3)$ & $(3.78, 10.1)$ & $(3.72, 10.2)$\\
		Unemployment Rate & $3.49$ & $3.35$ & $3.45$ & $(2.05, 6.56)$ & $(   2, 6.26)$ & $(2.03, 6.55)$\\
		Velocity & $3.62$ & $ 3.7$ & $3.73$ & $(2.26, 6.94)$ & $(2.28, 7.22)$ & $(2.29, 7.06)$\\
		Wages & $1.84$ & $1.84$ & $1.81$ & $(1.25, 2.91)$ & $(1.25, 2.95)$ & $(1.24, 2.88)$\\
		\bottomrule
	\end{tabular}
\end{table}

\subsection{Sampling Efficiency}

RNE of all chains is reported in Table~\ref{tab:app-eff}.
Parallels can be found to Table~\ref{tab:numbers}: AA is most efficient compared to SA for large posterior medians.
Although other parts of the Gibbs sampler also affect the figures, the median $\nu$ seems to be a strong factor in the ratio between the RNE's of AA and SA.
In both columns, the lowest value belongs to Interest Rate, and one of the largest to Stock Prices, respectively.
Moreover, the two columns predominantly stay close in magnitude.

\begin{table}[!t]
	
	\caption{RNE for estimating $\nu$ using AA, SA, and ASIS, along with the ratio between AA and SA and the mean of the three corresponding posterior median estimates.}
	\label{tab:app-eff}
	\centering
	\begin{tabular}[t]{lccccl}
		\toprule
		 & \multicolumn{4}{c}{RNE (\%)} & \\
		 & AA & SA & ASIS & AA / SA & Median\\
		\midrule
		CPI & 11.1 & 4.7 & 13.2 & 2.4 & $\sim2.18$\\
		Employment & 10.6 & 2.8 & 10.3 & 3.7 & $\sim2.31$\\
		GNP Deflator & 14.6 & 8.0 & 19.9 & 1.8 & $\sim2.21$\\
		Industrial Production & 12.7 & 3.2 & 12.6 & 3.9 & $\sim2.86$\\
		Interest Rate & 7.7 & 6.6 & 7.9 & 1.2 & $\sim1.22$\\
		\addlinespace
		Money Stock & 20.3 & 4.4 & 23.9 & 4.6 & $\sim3.58$\\
		Nominal GNP & 13.8 & 7.3 & 18.0 & 1.9 & $\sim2.32$\\
		Real GNP & 13.8 & 4.3 & 13.1 & 3.2 & $\sim3.05$\\
		Real per Capita GNP & 10.7 & 4.0 & 11.6 & 2.7 & $\sim2.84$\\
		Real Wages & 27.6 & 4.5 & 31.5 & 6.1 & $\sim 5.4$\\
		\addlinespace
		Stock Prices & 35.4 & 5.8 & 40.4 & 6.1 & $\sim6.03$\\
		Unemployment Rate & 14.7 & 4.2 & 14.0 & 3.5 & $\sim3.43$\\
		Velocity & 14.9 & 3.0 & 14.4 & 5.0 & $\sim3.69$\\
		Wages & 9.3 & 4.6 & 11.9 & 2.0 & $\sim1.83$\\
		\bottomrule
	\end{tabular}
\end{table}

\section{Discussion}\label{sec:con}

In this paper, efficient Bayesian estimation of the degrees of freedom parameter for the Student-$t$ distribution is investigated.
Specifically, the mixing speed of AA and SA is compared through various methods: an estimate for the expected augmented Fisher information, an extensive simulation study, and a suitable application.
AA is found to be beneficial both for real-world data and in most synthetic setups.
However, for extreme starting values, AA is found to be completely unreliable.

As a solution, ASIS is proposed to combine the good parts of AA and SA.
It is found to be as efficient as the better of AA and SA and also invulnerable compared to AA.
Moreover, ASIS is easy to implement as soon as AA and SA have been derived.
%It is impossible to emphasize enough, how highly the author thinks of the usefulness of this method in any setup.

Lastly, it is to be highlighted that the prime interest of the present work is the simulation-based description of the mixing speed of algorithms AA, SA, and ASIS.
In particular, the results are helpful for understanding the effect of re-parameterization on sampling efficiency, which is independent from execution time.
Therefore, optimization for effective sampling rate, which is a runtime-adjusted variant of RNE, and, granted, which is of most relevance to end users but is highly dependent on many factors~\parencite{Kriegel2017}, is not considered in the paper at hand.
One would almost certainly turn away from rejection sampling nowadays and rather experiment with other modern methods.
This remains for future research.

\section*{Acknowledgement}

We would like to thank Gregor Kastner for useful suggestions and helpful comments.

\printbibliography
\end{document}